# GIZA PYRAMIDS AND TAURUS CONSTELLATION

Vladan Panković , Milan  Mrdjen, Miodrag Krmar

Department of Physics, Faculty of Sciences, 21000 Novi Sad, Trg Dositeja Obradovića 4. , Serbia, vladan.pankovic@df.uns.ac.rs

**Abstract**

In this work we apply generalized A. Sparavigna method (use of freely available softwares (programs), e.g. http://www.sollumis.com/ , http://suncalc.net/#/44.557,22.0265,13/2014.12.29/09:22 , http://universimmedia.pagesperso-orange.fr/geo/loc.htm, http://www.spectralcalc.com/solar_calculator/solar_position.php and http://www.fourmilab.ch/cgi-bin/Yourhorizon) for analysis of possible astronomical characteristics of three remarkable Giza, i.e. Cheops, Chephren and Mikerin pyramids. Concretely, we use mentioned programs for determination of the Giza plateau longitude and latitude, moments of the sunrise and sunset for any day at the Giza plateau, and, simulation of the sky horizon above Giza plateau in any moment of any day, respectively. In this way we obtain a series of the figures which unambiguously imply the following original results. Any of three remarkable Giza pyramids (Cheops, Chephren and Mikerin) holds only one characteristic edge between apex and north-west vertex of the base so that sunrise direction overlap almost exactly this edge during 28. October. (There is a small declination of the overlap date by Chephren pyramid by which overlap date is approximately 23. or 24. October.)  Simultaneously, in the sunset moment for the same day, Taurus constellation (corresponding to holly bull in ancient Egypt mythology) appears at a point at the very eastern boundary of the sky horizon.

In this work we shall apply generalized A. Sparavigna method (she used freely available software (programs), especially http://www.sollumis.com/ , for simulation of the local Sun radiation direction, sunrise and sunset especially [1], [2], [3] , while, for the same aim, we use  http://suncalc.net/#/44.557,22.0265,13/2014.12.29/09:22 program too) for analysis of possible astronomical characteristics of three remarkable Giza pyramids, Kufu or Cheops great pyramid, Khafre or Chephren pyramid, and Menkaure or Mikerin pyramid. Additionally, for the same aim, we shall use freely available software http://universimmedia.pagesperso-orange.fr/geo/loc.htm, http://www.spectralcalc.com/solar_calculator/solar_position.php

and http://www.fourmilab.ch/cgi-bin/Yourhorizon.for determination of the Giza plateau longitude and latitude, moments of the sunrise and sunset for any day at the Giza plateau, and, simulation of the sky horizon over Giza plateau in any moment of any day, respectively.

Using http://www.sollumis.com/ and http://suncalc.net/#/44.557,22.0265,13/2014.12.29/09:22 we obtain a series of eleven simulated figures, Fig.1-11. These figures simply and clearly demonstrate the following. Any of three remarkable Giza pyramids, Cheops, Chephren and Mikerin, holds only one characteristic edge between apex and north-west vertex of the base so that sunrise direction overlap almost exactly this edge during 28. October (only for Chephren pyramid this overlap is little better for 24. than 28. October).

This 28. October, of course, does not represent any special day, solstice or equinox, for the sun motion. Nevertheless, we can check what happen during this day at the night sky above Giza plateau.

For this reason, using http://universimmedia.pagesperso-orange.fr/geo/loc.htm, we determine (in a satisfactory approximation) the Giza latitude 30.01°, Giza longitude 31.21°. Introduction of these data in http://www.spectralcalc.com/solar_calculator/solar_position.php for 28.October, we obtain 6h 06' sunrise moment and 17h 12' sunset moment.

Further, introduction of mentioned sunset data in http://www.fourmilab.ch/cgi-bin/Yourhorizon yields the following simulation of the night sky over Giza plateau in sunset time moment for 28. October presented at figures 12 and 13. At theses figures, as it is not had to see, Taurus constellation appears at a point at the very eastern boundary of the sky horizon. Later Taurus constellation moves toward west what is demonstrated at figures 14 and 15 (five minutes later), and figures 16 and 17 (ten minutes later).

In this way we obtain an original result representing an unambiguous correspondence between analogous geometrical characteristics of all three remarkable Giza pyramids and strictly determined event of the Taurus constellation appearance at the night sky.

All this, in common with well-known fact that Taurus constellation was connected with holly bull in ancient Egypt mythology, opens a real possibility that three remarkable Giza pyramids was ancient observatories for Taurus constellation.

In conclusion we can only repeat and point out the following. In this work we apply extended A. Sparavigna method (which used freely available software (programs) for simulation of the local Sun radiation direction) for analysis of possible astronomical characteristics of remarkable Giza pyramids. In this way we obtain a series of the figures which unambiguously imply the following original results. Any of three Giza pyramids holds only one characteristic edge between apex and north-west vertex of the base so that sunrise direction overlap exactly this edge during 28. October. Simultaneously, in the sunset moment for the same day, Taurus constellation appears at a point at the very eastern boundary of the sky horizon.

**Figures**

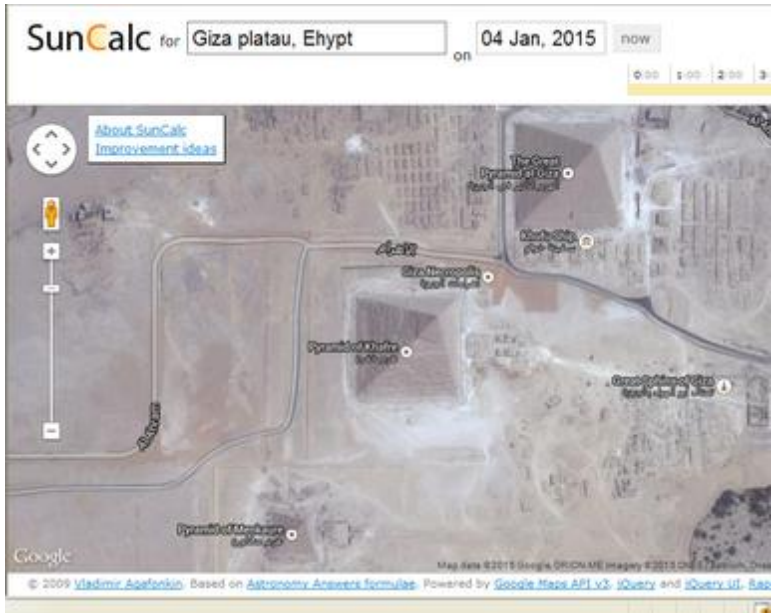

**Fig. 1 – Giza plateau with Cheops, Chephren and Mikerin pyramid**

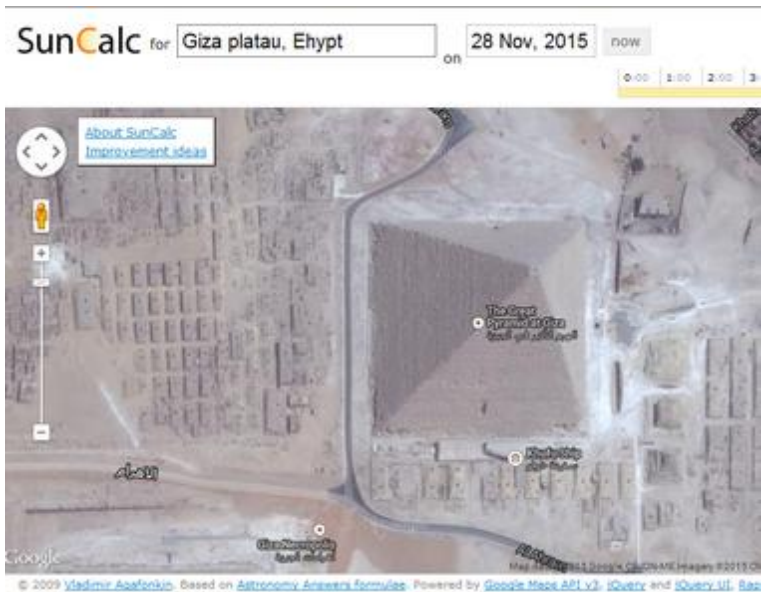

**Fig. 2 –Cheops pyramid**

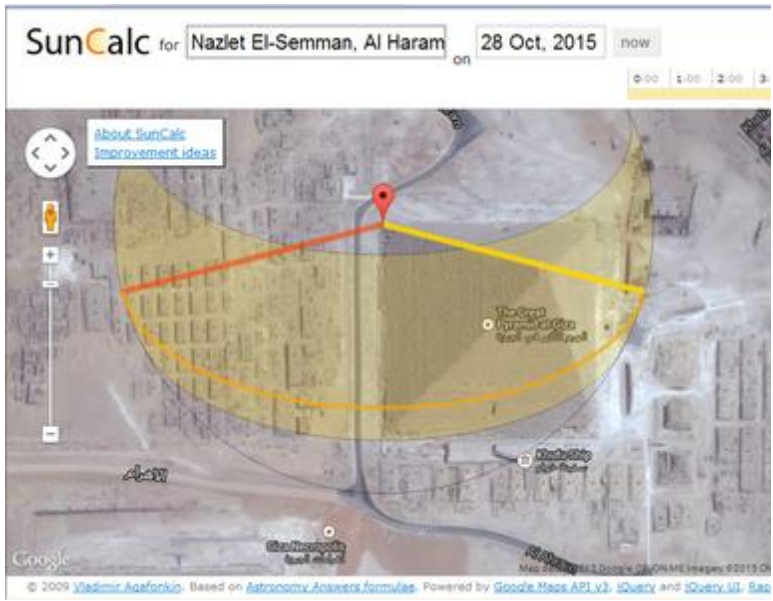

**Fig. 3 –Cheops pyramid with overlapped characteristic apex-basis edge and sunrise direction for 28. October**

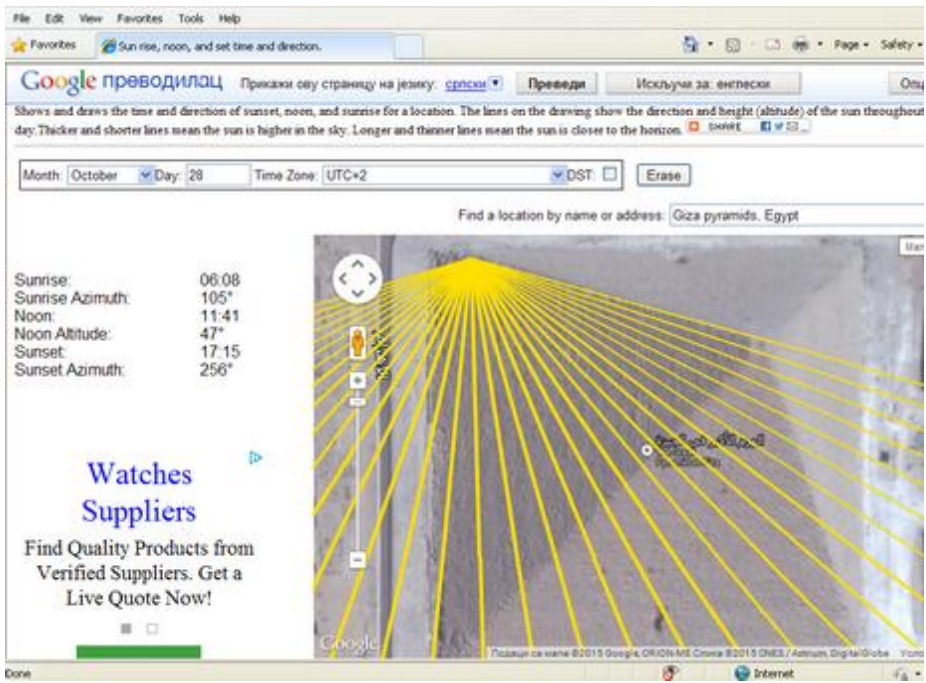

**Fig. 4 –Cheops pyramid with overlapped characteristic apex-basis edge and sunrise direction for 28. October**

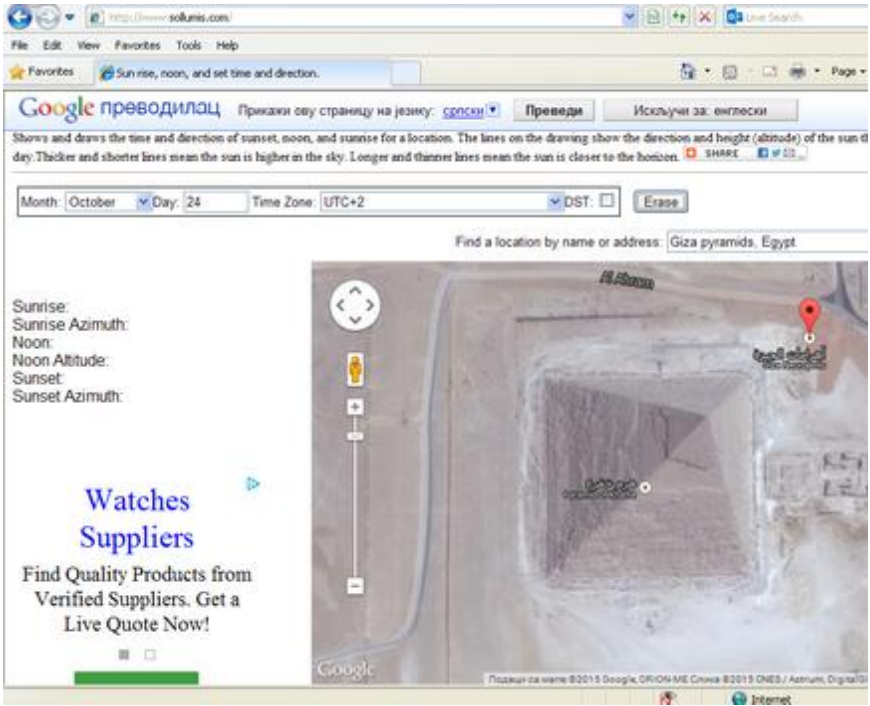

**Fig. 5 – Chephren pyramid**

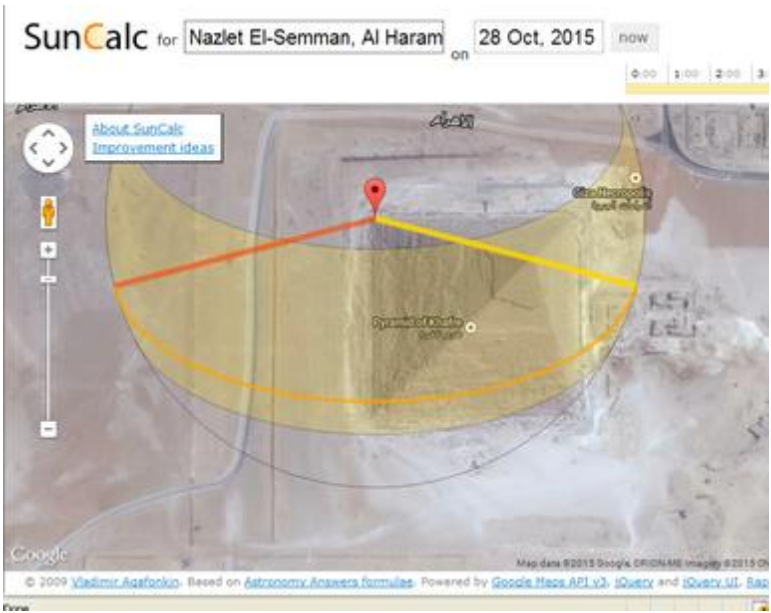

**Fig. 6 – Chephren pyramid with overlapped characteristic apex-basis edge and sunrise direction for 28. October**

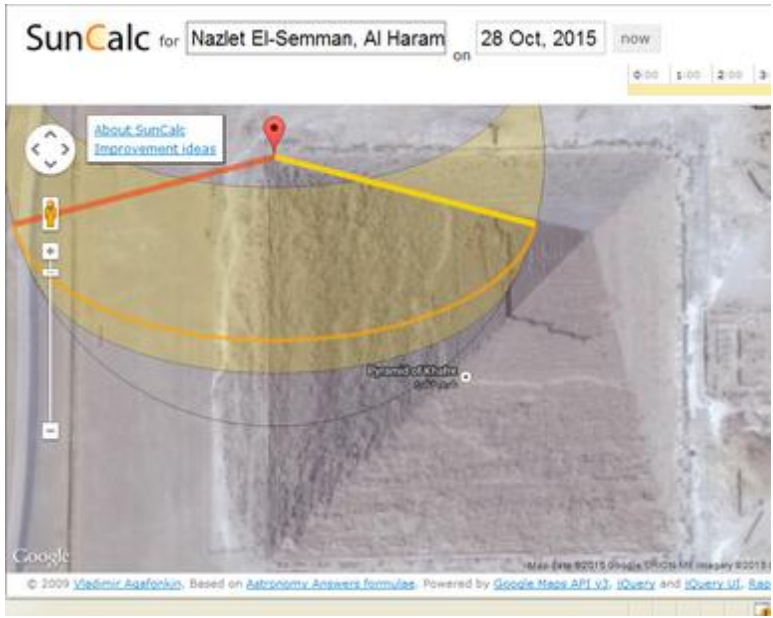

**Fig. 7 – Chephren pyramid with overlapped characteristic apex-basis edge and sunrise direction for 28. October**

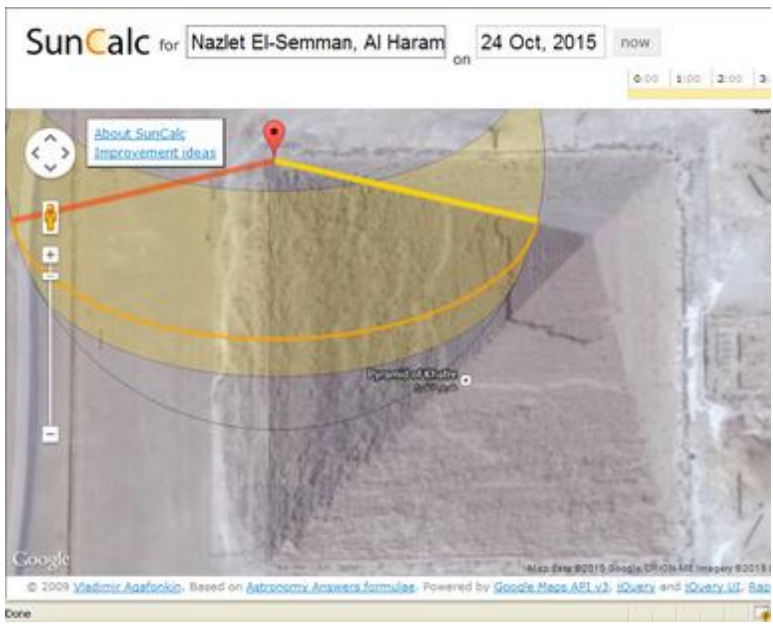

**Fig. 8 – Chephren pyramid with overlapped characteristic apex-basis edge and sunrise direction for 24. October**

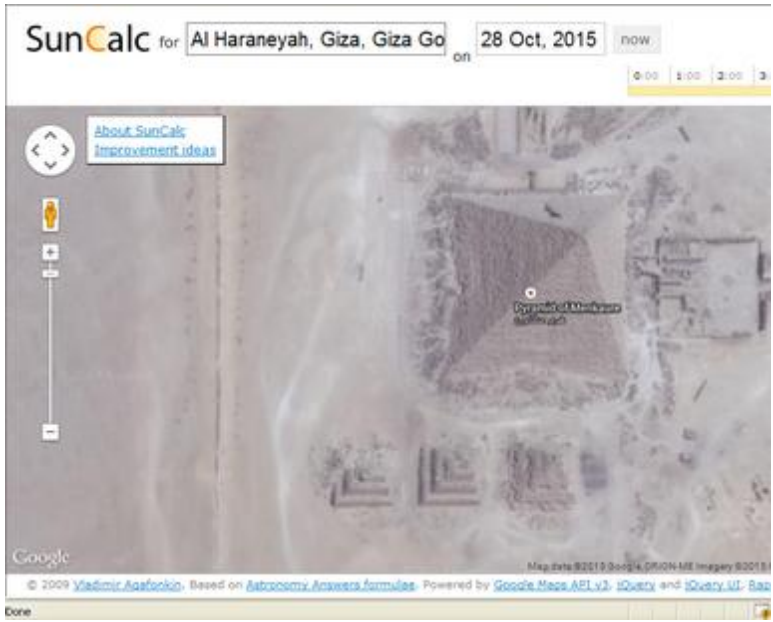

**Fig. 9 – Mikerin pyramid**

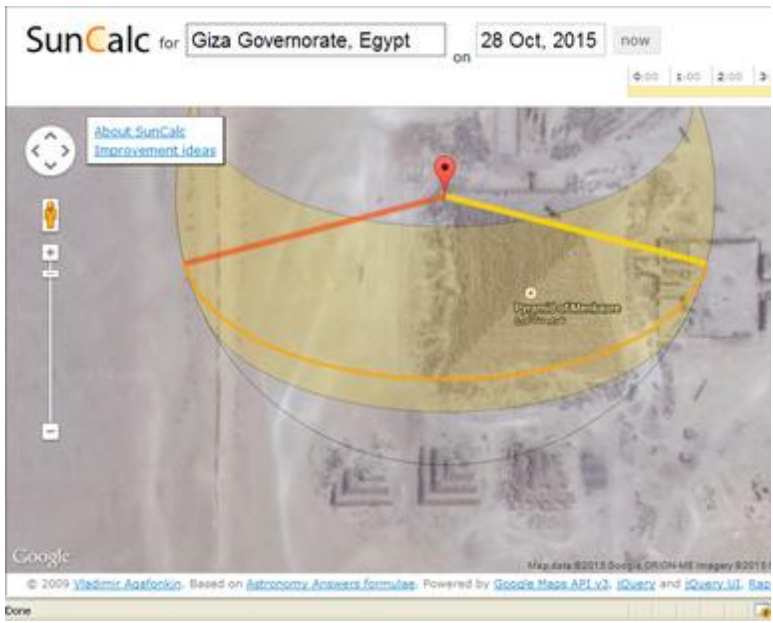

**Fig. 10 – Mikerin pyramid with overlapped characteristic apex-basis edge and sunrise direction for 28. October**

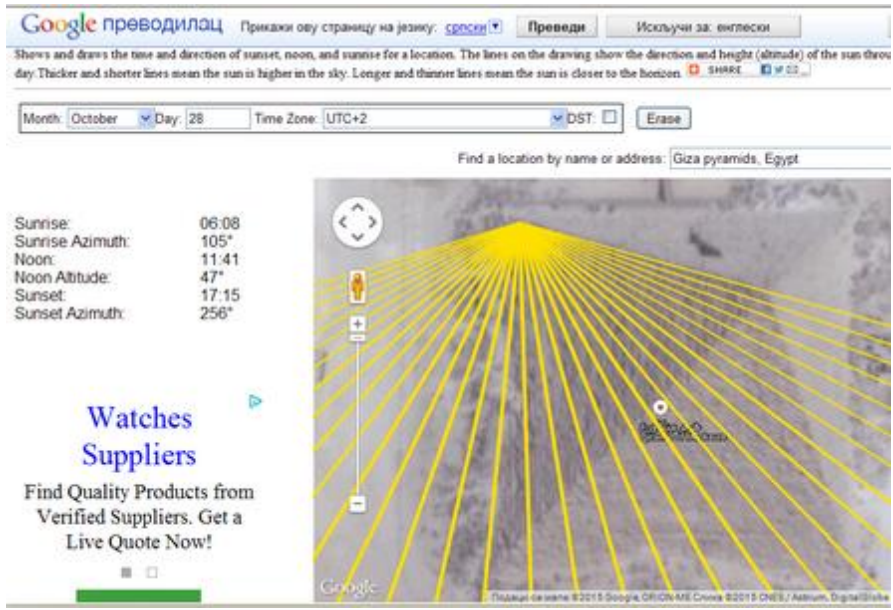

**Fig. 11 – Mikerin pyramid with overlapped characteristic apex-basis edge and sunrise direction for 28. October**

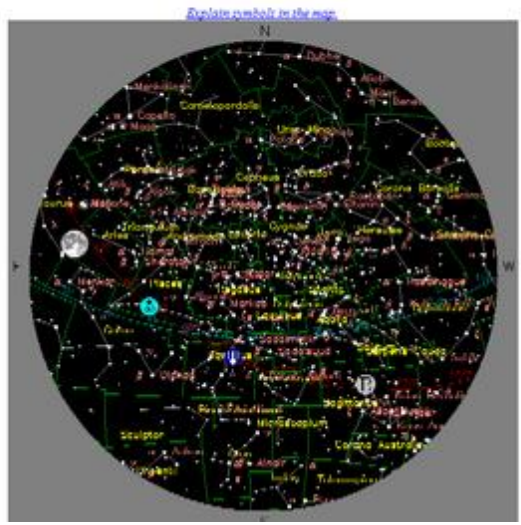

**Fig. 12 – Sky above Giza at 28. October in the sunset moment**

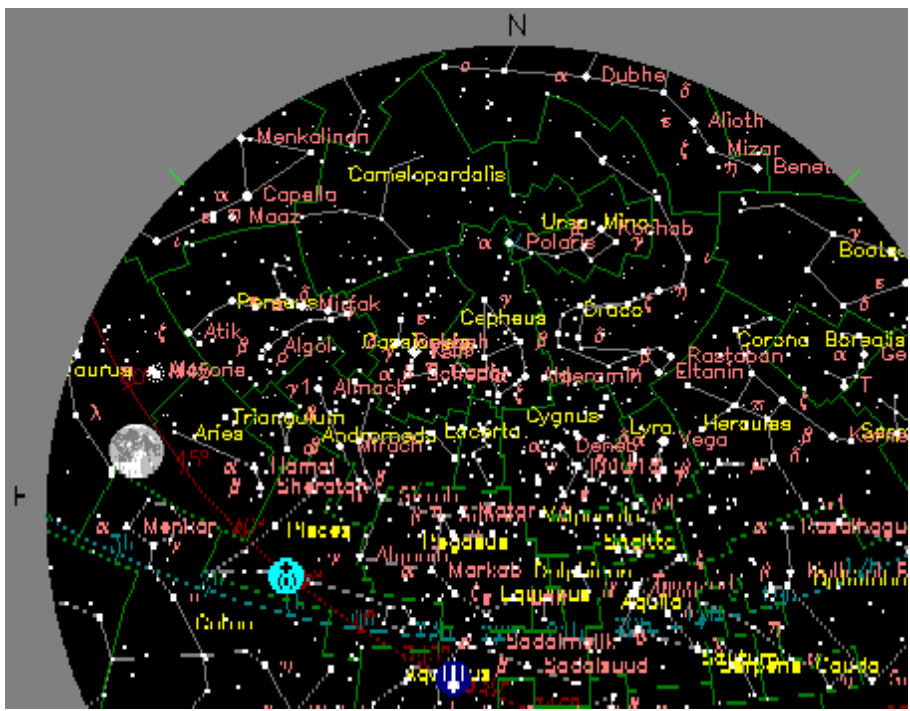

**Fig. 13 – Sky above Giza at 28. October in the sunset moment**

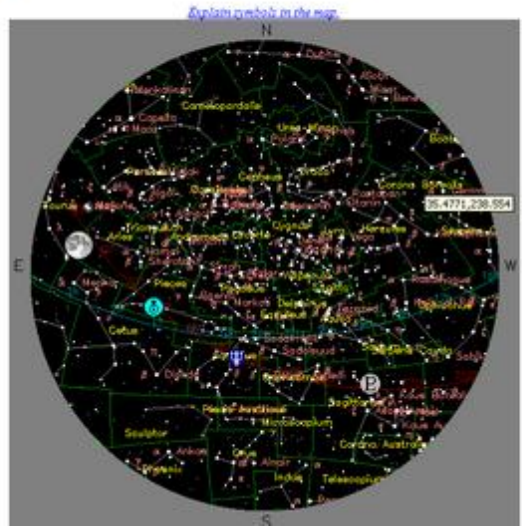

**Fig. 14 – Sky above Giza at 28. October five minutes after the sunset moment**

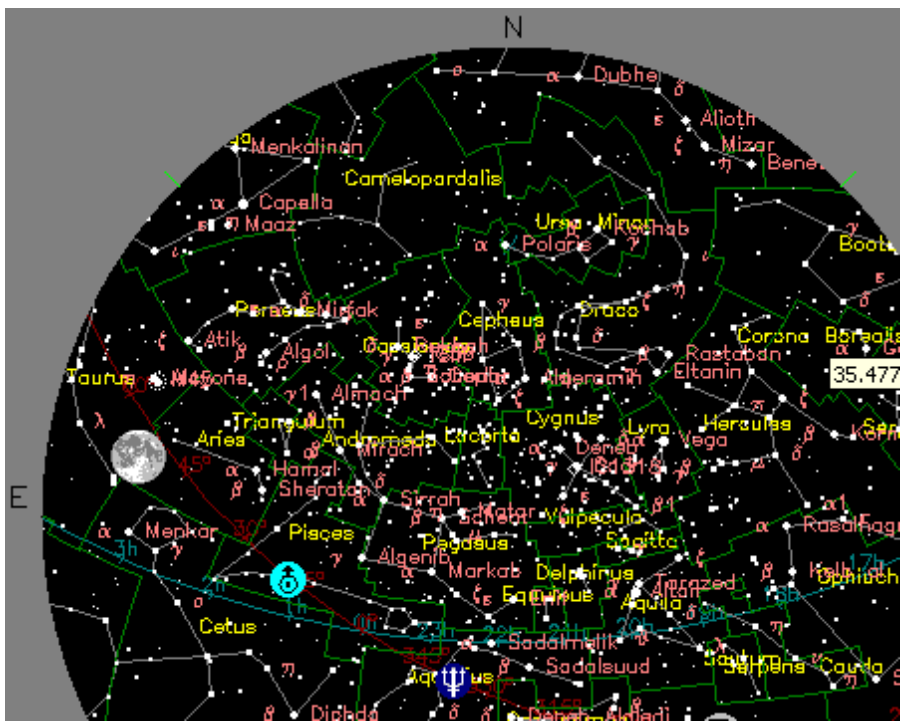

**Fig. 15 – Sky above Giza at 28. October five minutes after the sunset moment**

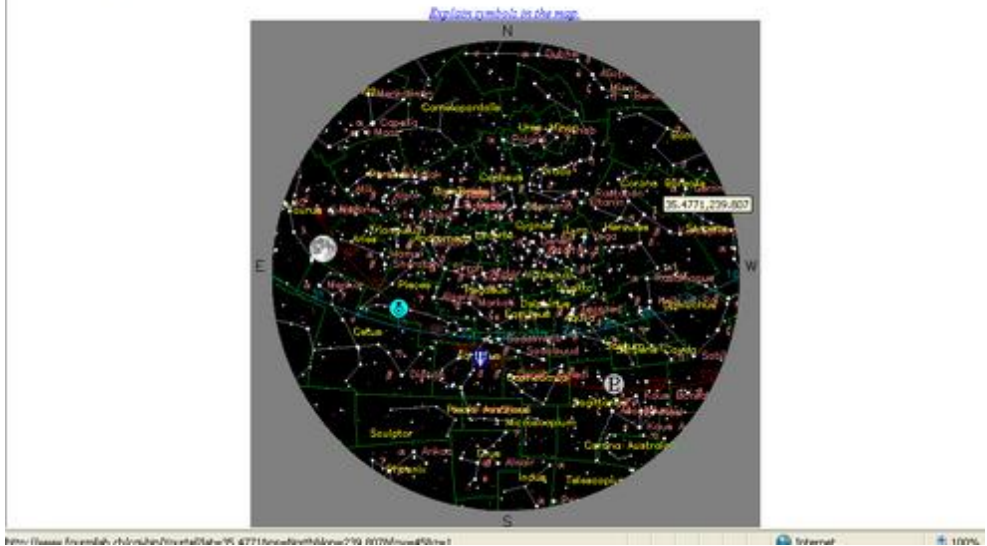

**Fig. 16 – Sky above Giza at 28. October ten minutes after the sunset moment**

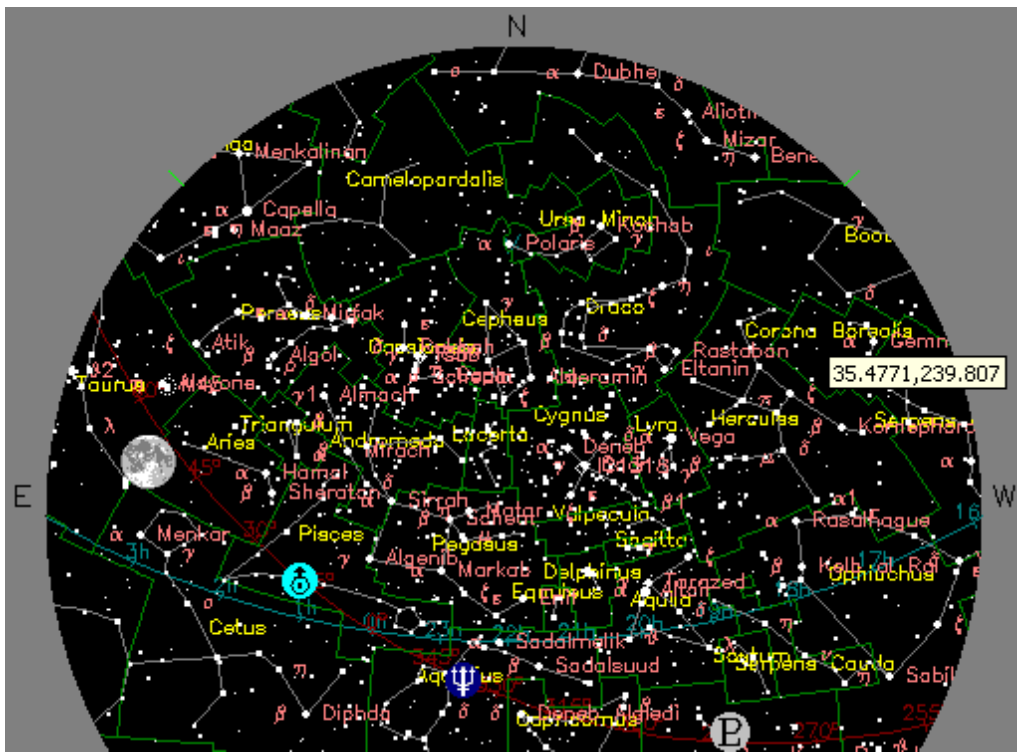

**Fig. 17 – Sky above Giza at 28. October ten minutes after the sunset moment**